\documentstyle{article}
\def\part{\partial}

\def\be{\begin{equation}}
\def\ee{\end{equation}}
\def\eq{\be}
\def\en{\ee}

\input epsf
\begin{document}
\addtolength{\topmargin}{-0.5in}
\addtolength{\textheight}{0.5in}
\addtolength{\oddsidemargin}{-.4in}
\addtolength{\evensidemargin}{-.4in}
\addtolength{\textwidth}{1.0in}
\begin{center}
\today     \hfill    LBNL-38960\\
{}~           \hfill UCB-PTH-96/24\\
{}~           \hfill astro-ph/9606052 \\

\vskip.7in {\Large\bf
Open universes from bubbles:
\vskip.05in
an introduction and
update}
\vskip.3in
J. D. Cohn\footnote{jdc@asterix.lbl.gov}\\
{\it Department of Physics, University of California, and\\
Theoretical Physics Group,
Lawrence Berkeley National Laboratory\\
University of California
Berkeley, CA 94720}
\end{center}
\vskip.2in
\begin{center}
{\small\bf Abstract}
\end{center}
{\small \baselineskip 0.4cm 
~~~~~~~~This is an introduction to models of open universes originating from
bubbles, including a summary of recent theoretical results for
the power spectrum.  
}
\begin{center}
{\small\it To appear in the Proceedings of the
 XXXI$^{\rm th}$ Moriond meeting, ``Microwave Background
Anisotropies.''}
\vskip .2in
{\small\rm PACS numbers 98.80Bp, 98.80Cq }
\end{center}

\setcounter{footnote}{0}
\vskip .5in

A flat ($\Omega_{tot}=1$)
universe has long been considered a generic
prediction of inflationary \cite{guth} models.  It has
recently been demonstrated that inflation can also 
produce viable open ($\Omega <1$) universes \cite{BGT}.
As the
 data on $\Omega$ has not yet converged, further investigation
of these models is worthwhile.
In the following, why and how bubbles give open universes
is reviewed, the main features of the new models
are outlined and current results and questions are
summarized.  Due to space limitations,
this is
necessarily just an overview, and readers are referred
to the cited papers and their references
for more in depth discussion and comprehensive referencing.

The basic reason bubbles give an open universe
can be illustrated with
empty Minkowski space, which
does not have a unique
coordinate system leading to a metric of the form
\eq
ds^2 = d\eta^2 - a^2(\eta) d \sigma^2 \; .
\label{frw}
\en
In empty space ($\rho = 0$),
with zero cosmological constant ($\Lambda = 0$),
the curvature $k$ obeys
\eq
H^2 =
\left({\dot{a}(\eta) \over a(\eta)}\right)^2 = {-k \over a^2(\eta)} \; .
\en

Consider the two coordinate systems for Minkowski space shown in
figure one.
\begin{figure}[b]
\begin{center}
\leavevmode
\epsfysize=1.2in \epsfbox{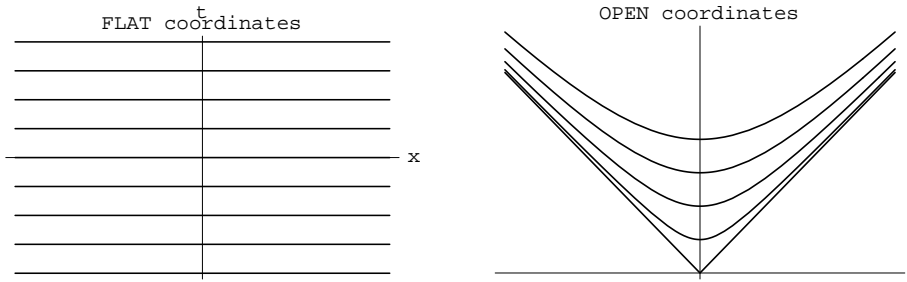}
\end{center}
\caption{Lines of constant time in
two coordinate systems for Minkowski space}
\label{fig:coords}
\end{figure}
On the left, lines of constant time are shown for a 
coordinate system with metric ($x^2 = x_1^2 + x_2^2 + x_3^2$)
\eq
ds^2 = dt^2 - dx^2 \; ,
\en
so one can read off that $a(t) = 1$.
As a result, for this coordinate system,
\eq
\left( {\dot{a}(t) \over a(t)} \right)^2 = 0 = -k \; ,
\en
the universe has $k=0$ and is therefore spatially flat.

A second coordinate system $(T,\chi)$ is shown on the right.  Here, lines of
constant time $T$ are shown.  In terms of
$(x,t)$, we have $
T = \sqrt{t^2-x^2} \; , \; \; \tanh \chi = {x/ t}
$.  The metric becomes
\eq
ds^2 = dT^2 - T^2 d\chi^2 \; .
\en
So in this case, we identify $a(T) = T$.  Then,
\eq
\left({\dot{a}(T) \over a(T)}\right)^2 = {1\over T^2} = - {k\over T^2}
\; \Rightarrow k = -1 \; ,
\en
a spatially open universe.

In empty space, there is no reason to prefer either coordinate system.
However,
if there is matter present, one or the other might be preferred. 
For example,
if the background matter distribution is homogeneous
in space only in a particular coordinate system,
the metric has the form eqn.(\ref{frw}) only for
that choice of coordinates.
A bubble is said to create an open universe because inside
it
the $(T,\chi)$ coordinate
system is preferred.

To get a bubble,
start with a system (here described by a field $\phi$)
stuck in a false vacuum.
A bubble forms when
a region of space, the bubble interior,
tunnels to the true vacuum.
\begin{figure}[t]
\begin{center}
\leavevmode
\epsfysize=1.7in \epsfbox{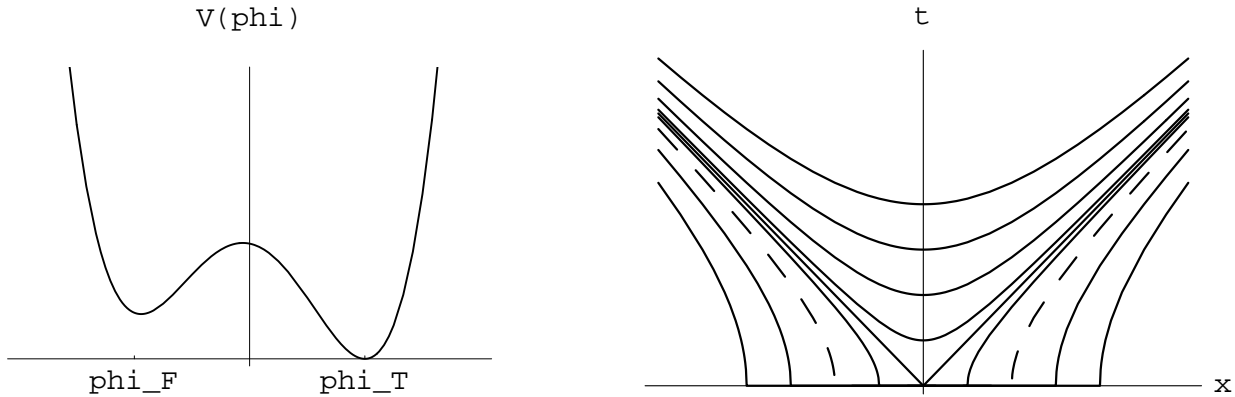}
\end{center}
\caption{On the left, a potential with a false and true vacuum.
On the right, lines of constant $\phi$ after the bubble
nucleates.}
\label{fig:poten}
\end{figure}
An example of a potential where this could happen is
shown on the left of figure two.
After tunneling, there is a true vacuum interior, where
$\phi = \phi_T$, a bubble (domain)
wall, and a false vacuum exterior where $\phi = \phi_F$.  
 The lowest energy solution for tunneling in
Minkowski space \cite{col}
has spherical $O(4)$ symmetry, $\phi = \phi(x^2-t^2)$.
As the matter (the field $\phi$) is constant on surfaces of constant 
$t^2-x^2= T^2$, it is natural to use the open coordinates \cite{go-co},
where the matter is homogeneous in space.

On the right side of figure two is a spacetime diagram showing
lines of constant $\phi$.
The dashed line is the bubble wall, and the
horizontal axis is the nucleation time.
As a function of the flat coordinate time $t$, the bubble wall
moves out. Classically the energy difference between
the false and true vacua (the latent heat) goes into
accelerating 
 the
wall outwards. 
Inside the future light cone of the center of the bubble is
a patch with open coordinates.  As the light cone is
$T=0$ in this coordinate system, and the bubble wall
in this example is exterior to the light cone (approaching
it at large times), the bubble wall is ``before''
$T=0$.

This picture can be used to generate a candidate
for the early universe by making some modifications.
Gravity must be
included, the field $\phi$ is taken to be part of the
inflationary potential,
and the bubble nucleates in a vacuum, the de Sitter vacuum,
$\phi \sim const.$  Nucleating a bubble in de Sitter space
also means that the global structure of spacetime is more
complicated--because the spacetime is expanding in the false 
vacuum, the future light cone of the bubble does not eventually
cover the whole future of the space (see, e.g. \cite{ballen}).
Many of the calculations require normalizing on a Cauchy
surface (a spacelike hypersurface
which every non-spacelike curve intersects exactly
once \cite{cauchy}) and thus use information exterior to
the open universe.

The
interior of the vacuum bubble has energy density depending on
the true vacuum potential, $V(\phi_T)$.  If
$V(\phi_T) \approx 0$, then after $\phi$ has tunneled,
the universe is effectively empty.  All the energy density
is in the bubble wall and not in the resulting open universe,
which has $\Omega \sim 0$ and stays that way.  Under
our assumptions, to 
get a universe with something in it means that
$V(\phi_T) \ne 0$.
If $V(\phi_T) \ne 0$, then the resulting vacuum is also an
expanding de Sitter space, that is, it is still inflating.
So one needs a bubble plus inflation before and after nucleation.
If one has inflation but no bubble, the amount of inflation providing
sufficient homogeneity to agree with observation drives 
$\Omega \rightarrow 1$ (for a nice summary in the
context of bubbles, see \cite{juantalk}).

Bucher, Goldhaber and Turok \cite{BGT}
combined bubble formation with inflation to give an open universe.
They use four stages, corresponding to potential features 
sketched in figure 3:
\begin{figure}[t]
\begin{center}
\leavevmode
\epsfysize=2.5in \epsfbox{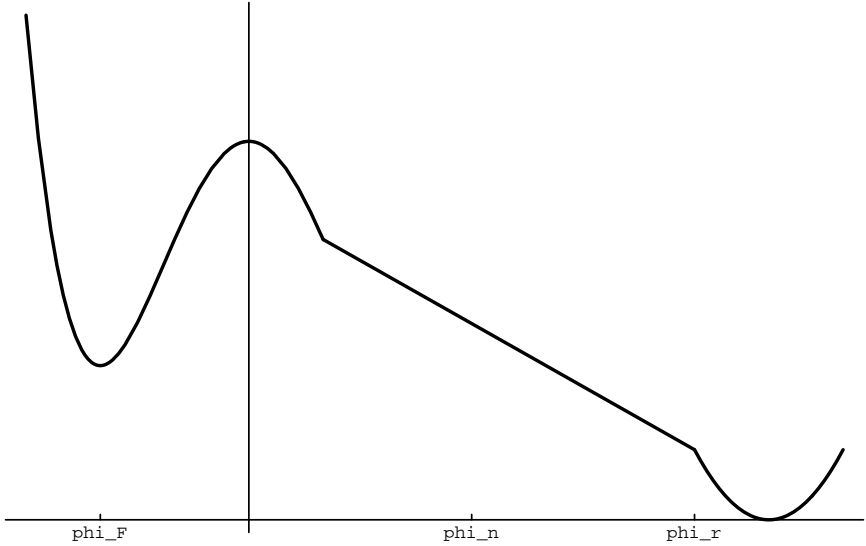}
\end{center}
\caption{A potential with the features
that give a viable open universe inflationary model.}
\label{fig:nucle}
\end{figure}
\begin{itemize}
\item  First, on the far left,
the system is trapped in a potential well (at
$\phi =\phi_F$),
a false vacuum.
Inflation occurs, driving the system to the Bunch-Davies vacuum
(the natural vacuum here, respecting the symmetries and
going over at short distances
to the Minkowski space vacuum)
and wiping out inhomogeneities.
\item Then $\phi$ tunnels to
$\phi \sim \phi_n$, resetting $\Omega \sim 0$;
\item  Next, on the approximately linear part of
the potential, 
from $\phi \sim \phi_n$ to
$\phi \sim \phi_r$,
slow roll inflation drives $\Omega$ up to
a specific value less than or equal to one.
\item  Finally, on the far right, at $\phi \sim \phi_r$,
$\phi$ decays, leading to the usual reheating after inflation.
\end{itemize}
The endpoint of the tunneling $\phi \sim \phi_n$
 will be called the true vacuum (since $\phi \sim const.$), even
though it is also the starting point of the slow roll inflationary
stage.

There are constraints on the potential so that this picture is
consistent.
The potential $V(\phi)$ has to be such that the tunneling rate is
rare, so that tunneling from $\phi \sim \phi_F$ only occurs after
a long period
of inflation has erased initial conditions.
The false vacuum well cannot be too flat or $\phi$ will
stochastically go over the barrier rather than tunneling through
it (the Hawking-Moss instanton) \cite{hawkmoss}, leading to
density perturbations which are too 
large \cite{linde-book}.
The potential $V(\phi)$ must be tuned to
within a few percent so that at the end of the slow roll stage
$0.1 \le \Omega \le 0.9$.

A simple polynomial potential (up to order $\phi^4$) does not
satisfy these conditions \cite{lm}, although in principle an
effective potential could arise in a supergravity theory
with the right features.  One can also use more fields,
driving inflation by one field while
the other one tunnels.  
Some of the two-field variants of the model lose 
the ability to predict a definite value of
$\Omega$ \cite{linde}, and lead to more questions
which have been studied extensively but are outside
the scope of this summary.  

The density perturbations in these
models are a modification of those in $\Omega = 1$
inflationary models.  For $\Omega =1$ inflationary
models, inflation
drives the universe to the vacuum, and the quantum
fluctuations
of the inflaton in this vacuum become seeds for structure
formation once they re-enter the horizon.
In these open universe bubble models, 
there are vacuum fluctuations (from the false vacuum
before tunneling) 
outside the bubble wall.  
These evolve through
the bubble wall, a time dependent background, to
reach the open universe.  There is a continuum of 
these modes
and sometimes (depending on properties of
the wall and the mass in the
false vacuum) a discrete mode.
The bubble wall itself also has fluctuations.
The calculations incorporate methods for
bubbles in Minkowski space \cite{minkp,hsty}, and
techniques for including gravity as in \cite{TS};
see these papers 
and references therein for background.  
Various choices for the
true and false vacuum masses and the wall profile
have been considered.
 
Calculations so far have been for vacuum fluctuations due to one
field,
in two field models the second field has been frozen out.
Thus one  solves 
\eq
\left(\Box -
{\delta^2 V(\phi) \over \delta \phi^2}|_{\phi = \phi_{\rm bkgd}}\right)
\delta \phi = 0
\en
The gauge invariant
gravitational potential produced by the continuous modes
has been found assuming
a constant value of $H$ in the resulting de Sitter space.
The power spectrum has been calculated for
a thin wall with arbitrary false vacuum mass $M^2 \ge 2 H^2$ and zero
radius \cite{BT} and
nonzero radius \cite{new,jdc}.  It has also been
found for 
a large false vacuum mass and varying wall profile
(with the restriction that the wall is completely
exterior to the light cone of the bubble center, the open
universe) \cite{new}.  In all of these cases, the contribution
to the fluctuations is within an envelope between
$\coth \pi p/2$ and
$\tanh \pi p/2$ times a scale invariant spectrum.
(The correspondence with wavelength
is $p = k/(H_0\sqrt{1-\Omega_0})$.)
Thus, except at very large scales, where cosmic variance
interferes with their measurement,
these spectra all coincide. 

In open de Sitter spacetime, when there
is no bubble present, the Bunch Davies vacuum
fluctuations of a field of mass $M^2 < 2 H^2$ has a continous
spectrum of fluctuations plus a discrete mode \cite{vacmode}.  This
discrete mode was shown to be part of
the complete basis of states \cite{vacmode} and
is normalizable on a Cauchy surface, but
not inside the open universe (which does not contain a 
Cauchy surface). 
In the presence of a bubble, this discrete `supercurvature'
mode with $-1 \le p^2 \le 0$ may remain, depending upon the false
vacuum mass and the wall profile. 

If the field providing density perturbations
does not change its mass across the wall (as can happen in two
field models), and has $M^2 < 2 H^2$, this mode appears \cite{cmb,BY}.
Its contribution to density fluctuations has been found and used to
constrain models when combined with the CMB for various cases
 \cite{cmb,BY,juan,twosur}.  It does not seem to give a very
strong constraint.  However, 
it has been argued that this state's
contribution can
be enhanced significantly when the ratio of false to true vacuum
energy is large,
ruling out some models \cite{twosur}.  There
is not yet complete agreement on this.
If there is a mass change across the wall \cite{new}, 
an analogue of this state may persist.  When matched
across the bubble wall, the
original vacuum 
discrete mode does not automatically remain normalizable,
but a mode with
some other value of $-1 \le p^2 <0$ may become
normalizable instead.
The effects of this new mode 
have been found in a two field model for a thin wall \cite{new}.
This mode disappears if
the false vacuum mass is large enough (as argued
in \cite{new}, the false vacuum mass is necessarily large for most
one field models considered so far, so that it does not
appear there).

There are also fluctuations of the wall itself, 
contributing mostly at large scales, and
considered in \cite{gv,lm,hsty,gar96,juan,new}.   
Requiring these fluctuations to be small constrains models. 
Schematically, the allowed models have
a high barrier between the false and true vacuum,
producing
a large surface tension and making it energetically
unfavorable for the wall to fluctuate.

In summary, the continuum modes reproduce a scale invariant
spectrum starting at very large scales.  The vacuum supercurvature
mode or its analogue (when present) 
and the wall fluctuation (also a supercurvature mode)
 have been
used so far
to constrain possible potentials by requiring that their
contribution be small.

Many fits to the data \cite{data} have been made for open universes
assuming a scale invariant \cite{conformal} spectrum, and,
except at large scales, the models considered so far also
have this spectrum, and so carry over.
Many things remain to be done, including adding gravity
waves and tilt, and motivating models from particle theory. 

Acknowledgement:
I thank F. Bouchet for the chance to speak and for organizing the
conference, and the participants (D. Bond, P. Ferreira, A. Liddle, 
N. Turok
and M. White in particular) for conversations.  I also
thank K. Benson, M. Bucher, and V. Periwal
 for conversations and D. Scott and
M. White for comments on the draft.
This work
was supported in part by the Director, Office of
Energy Research, Office of High Energy and Nuclear Physics, Division
of High Energy Physics of the U.S. Department of Energy under Contract
DE-AC03-76SF00098 and in part by the National Science Foundation under
grant PHY-90-21139.

\end{document}